# A Novel Algorithm for Exact Concave Hull Extraction

Kevin VanHorn[1*], Murat Can Çobanoğlu[1]

[1]Lyda Hill Department of Bioinformatics, University of Texas Southwestern Medical Center, Dallas, Texas, 75390

*Contact: kevin.vanhorn@utsouthwestern.edu

# Abstract

Region extraction is necessary in a wide range of applications, from object detection in autonomous driving to analysis of subcellular morphology in cell biology. There exist two main approaches: convex hull extraction, for which exact and efficient algorithms exist and concave hulls, which are better at capturing real-world shapes but do not have a single solution. Especially in the context of a uniform grid, concave hull algorithms are largely approximate, sacrificing region integrity for spatial and temporal efficiency. In this study, we present a novel algorithm that can provide vertex-minimized concave hulls with maximal (i.e. pixel-perfect) resolution and is tunable for speed-efficiency tradeoffs. Our method provides advantages in multiple downstream applications including data compression, retrieval, visualization, and analysis. To demonstrate the practical utility of our approach, we focus on image compression. We demonstrate significant improvements through context-dependent compression on disparate regions within a single image (entropy encoding for noisy and predictive encoding for the structured regions). We show that these improvements range from biomedical images to natural images. Beyond image compression, our algorithm can be applied more broadly to aid in a wide range of practical applications for data retrieval, visualization, and analysis.

# Introduction

Accurate region extraction for imaging data is an essential but complicated task. Applications include object detection for autonomous vehicles[1], anomaly detection in commercial robotics[2], pathfinding and collision avoidance[3], region of interest extraction for hyperspectral satellite imagery[4], and analysis of subcellular morphology[5]. Specifically, higher region precision is essential when there is no room for compromise in data integrity.

We present a fast and generalizable concave hull algorithm that computes pixel-perfect and vertex-minimized contours. This approach is termed Adaptive Geometric Contouring (AGC). As almost no real-world object is truly convex, this novel technique solves a generalizable problem that is widely applicable to fields including computational geometry, visualization, pattern recognition, and image processing. Furthermore, by providing a deterministic and exact contour, the solution is immediately beneficial to downstream analysis, including curvature detection and region compression or extraction. Unlike other concave hull algorithms, AGC represents regions with exact precision, providing immense benefits where data integrity is a priority.

One of the more recent efforts with the similar goal of extracting local representations is the Adaptive Particle Representation (APR) work by Cheeseman and colleagues[6]. In this effort, instead of defining objects with concave or convex hulls, regions are divided into regularly shaped but irregularly sized grids. The size of the grid controls the granularity of the representation. The problem with this, and other similar approaches such as quadtrees or octrees, is that real world objects are almost never bounded regularly. Therefore, there is inevitable mismatch between the representation and the nature of the object. AGC obviates this restriction by allowing arbitrary shape definitions.

Among the myriad applications of a contouring algorithm, we chose to reduce our ideas to practice in an area where data integrity can be a legal requirement depending on local jurisdiction: biomedical image data compression. Biomedical imaging is an evolving technology that provides researchers and healthcare professionals with insight that would otherwise be inaccessible. Improvements in this field can help us build better diagnostic tools, improve our understanding of biology, and perform better treatments[7]. Lossy compression and pre-processing techniques improve image accessibility but can lead to a loss in clinical precision and introduce reconstruction errors. Laws and regulations covering Diagnostically Acceptable Image Compression (DAIC) often vary per country and medical field[8]. Thus, the use of lossy compression on biomedical images is often quite complicated in practice since most jurisdictions have laws regulating data integrity.

Lossless methods provide perfect data reconstruction and are therefore particularly suitable for biomedical images. However, they tend to be slower and supply a compression ratio of 1.5:1 to 3:1 on average compared with lossy which can have ratios upwards of 20:1 without discernable loss in visual integrity[9]. In addition, most traditional lossless techniques are designed for natural, continuous tone images and are ill-suited for medical images which are often characterized by a high nominal bit depth and sparse distribution of intensity values[10]. In certain medical image modalities, including Computed Tomography (CT) and Magnetic Resonance (MR), universal compression has been shown to outperform lossless image compression algorithms[10]. Such a technique was integrated into the Keller Lab Block format, where the standard BZIP2[11] software was used to perform parallel lossless compression of 5D light-sheet microscopy data for its speed and efficiency[12].

In summary, many compression techniques exist that can be applied to biomedical images, but the performance of each method and specific tolerance of loss is highly dependent on the domain or dataset. Thus, it is necessary for a compression format to exist that can be fine-tuned to the content stored. Without accessible lossless image compression methods, researchers and medical professionals may choose to avoid compression entirely or integrate convenient but nonoptimal compression algorithms. We propose a method that addresses these concerns and enables a user to maintain data integrity on a case-by-case basis while still capitalizing on bioimaging characteristics such as sample sparsity.

Extending the AGC algorithm, we introduce the lossless Adaptive Geometric Contour Representation (AGCR) that provides benefits for both image compression, visualization, and retrieval (Figure 1). We designed AGCR for sparse images with a high bit-depth and/or high resolution, however, this method is easily applicable to any raster image. In a near-lossless configuration, AGCR outperforms state-of-the-art compression methods of these modalities through a context-dependent strategy. Fully lossless AGCR is comparable to, or better than, similar lossless strategies for biomedical images, proving to be effective for depth-based imagery, and consistently more effective than alternatives for raw 16-bit natural images.

Most notably, AGCR enables entirely novel applications such as the separation of multiple sample and background rasters. For example, with this technique one can archive a background with maximal lossless or even lossy compression, while storing the foreground separately to be retrieved with a faster lossless codec. This process is highly configurable and can be specifically tuned to a specific image or dataset. Furthermore, we support the use of multiple compression codecs for different regions of an image, which serves as the backbone of our adaptive approach. Finally, as a representation, AGCR enables the visualization of 2D contours. These provide a useful preview or a means with which to decode, transmit, or display compressed regions selectively.

Multiple geometric methods for lossy image construction exist that integrate triangulation algorithms for progressive representation. However, they do not enable the lossless representation of regions of interest and are incompatible with high bit-depth and high-resolution biomedical images. One such method, adaptive mesh representation[13], approximates the image mesh on reconstruction. Superfluous vertices are necessary to maintain the integrity of the mesh and, without intensive non-uniform sampling and reconstruction strategies across vertices, the algorithm fails to efficiently represent an image of high density. This technique has been applied to biomedical images but aims at improving restoration rather than functioning as a partitioning technique, thus it is not visually informative[14]. Adaptive graph-based solutions exist but are inflexible in their partitioning strategy[15–18] or only function effectively as a lossy solution[19].

Various configurable image-processing strategies exist that one can apply in the context of biological and medical imaging. Volume of interest (VOI) coding has been proven effective in telemedicine and enables the user to choose lossy compression in some rectangular areas and lossless in others[20,21]. These regions of significance are determined by diagnostically important regions. Such an approach can also be applied to data transfer and visualization through variable levels of detail[22,23]. Data elimination is another technique in this domain due to the sparse nature of many biomedical images[24–27]. Use of the Adaptive Geometric Contouring algorithm complements both VOI and data elimination in two ways. First, we enable multi-level partitions as opposed to binary assignments, with each partition allowing a configurable loss or compression method. Second, we supply pixel-perfect contours rather than bounding boxes, ensuring clinical integrity of relevant data in irregularly shaped regions. Finally, we store the intensities as offsets from a local region minimum alongside histogram packing thereby allowing flexible reduction in the effective number of bits used. The end effect is that AGCR enables lossy to lossless storage, compression, and/or network transmission.

# Results

## Adaptive Geometric Contouring Algorithm

Convex and concave hull algorithms aim to construct a polygon that envelops a set of points which respect to some objective function typically computing a minimum area for the bounding region[28]. When used as a partitioning structure, concave hull algorithms are ineffective at representing non-overlapping sets of pixels due to their approximate nature. Thus, we developed the Adaptive Geometric Contouring (AGC) algorithm that maintains region integrity through a pixel-perfect objective function that is vertex-minimized to improve spatial efficiency.

Given an image with *n* four-connected pixel regions, we construct concave hulls termed geometric contours that represent the outer contour of each region in the fewest number of representative vertices. To enable lossless reconstruction of every region in the image, we enforce the pixel-perfect constraint such that pixels outside of the containing four-connected region are never included in a contour. To represent a polygon with one or more interior boundaries, we compute the geometric contour of each contained shape and subtract each hole from the original polygon.

Acting on a four-connected region of pixels *R*, the Adaptive Geometric Contouring algorithm (*Algorithm 1*) functions by storing a start, previous, and current vertex labeled as $v_0$, $v_{i-1}$, and $v_i$ respectively. We define filled positions as pixels contained in *R* and empty positions as pixels not in *R*. We check each position counterclockwise about $v_i$ until the last empty and first filled vertices are found, $v_e$ and $v_f$. Two rays are cast in the direction of the vector from the $v_e$ and $v_f$ toward $v_i$. We halt if the $v_e$ ray encounters a vertex or if the $v_f$ ray exits the shape, at which point the longest edge is taken, variables are adjusted, and the process is repeated until the starting vertex is reached. When moving diagonally toward internal angles, we correct to the $v_f$ ray-pixel intersect. Finally, let $M = w \cdot h$ where $w, h$ are the width and height of the image.

{Precondition: Acting on a given region of vertices, $R$.}
```
 1: procedure AGC(R)
 2:    set v_0 to bottommost-left vertex of R
 3:    set v_i to v_0 and push v_0 to S
 4:    loop while v_i ≠ v_0:
 5:        set v_{i-1} to closest adjacent vertex from v_i in direction of v_i − v_{i-1}
 6:        set v_e and v_f as the last empty and first filled vertex about v_i counterclockwise
 7:            set v_i to v_0 − [1,0] and d to v_f − v_i
 8:        loop ℓ from 1 to M while v_i ≠ v_0:
 9:            if v_e + d · ℓ ∈ R:
10:                set v_{i-1} to v_i and v_i to d · ℓ + v_e
12:            else if v_f + d · ℓ ∉ R:
13:                set v_{i-1} to v_i and v_i to d · (ℓ − 1) + v_f
14:            else if d_x ≠ 0 and d_y ≠ 0:
15:                if {v_e + d · ℓ − [0, d_y] ∈ R and v_e + d · ℓ − [d_x, 0] ∈ R }
17:                or { v_e + d · ℓ + [d_x, 0] ∈ R and v_e + d · ℓ − [d_x, 0] ∈ R and v_e + d · (ℓ + 1) = v_0 }:
18:                    set v_{i-1} to v_i and v_i to d · (ℓ − 1) + v_f
18:                else continue loop
19:            push v_i to S and break loop
23:    return S
```
{Output: A shape represented by a list of vertices, $S$.}

***Algorithm 1:*** *Adaptive geometric contouring visits 4-connected components and wraps them with vertex-minimized concave shapes.*

To further optimize the AGC algorithm to produce polygons with less vertices, we can run a second pass on the resulting counter from *Algorithm 1*. This optimization phase visits each vertex to find the longest possible edge that does not include vertices out of the region. Although not necessary for the core algorithm to function, a higher level of optimization can offer improvements in applications like compression where spatial efficiency takes priority.

{Precondition: Acting on a given region of vertices, $R$ of size $N$.}

```
 1: procedure OptimizeRegion(R)
 2:    loop k to N − 1:
 3:        set v_0 to R_k and θ_{i−1} to 180
 4:        loop n from k + 2 to N − 1:
 5:            set v_i to R_{n−1} − v_0
 6:            set v_{i+1} to R_n − v_0
 7:            set θ_i to angle from v_i to v_{i+1}
 8:            set c to R_{n−1} − v_0 + nearest vertex counterclockwise from −v_i
 9:            set θ_{i+1} to angle from v_0 to c
10:            if v_{i+1}.x is 0 and v_{i+1}.y is 0 then break loop
11:            else if θ_i ≥ 0 and |θ_i| < |θ_{i+1}| and θ_i ≤ θ_{i−1}
12:                push vertex to S
13:                set θ_{i−1} to θ_i
14:            else break loop
```

{Output: An optimized list of vertices for this region, $S$.}

***Algorithm 2:*** *We further optimize geometric contours to address longer edges that were not minimized in the first stage of contouring.*

Adaptive Geometric Contouring Representation

AGCR is a fully lossless partitioning strategy that facilitates the compression, visualization, and retrieval of images with a high bit-depth, dynamic range, and/or resolution. Using the AGC algorithm, we define geometric contours as vertex-minimized concave hulls that contain every pixel in each region. This approach provides spatially efficient partitioning for irregular biological or other visual regions of interest.

By default, AGCR partitions an image with a probability-based thresholding approach. We inform the resulting *k*-level thresholding using the Gini coefficient[29] of the image to reflect its sparsity. Each 4-connected region of a thresholded image is defined by a geometric contour and the contained intensities are stored separately. The application automatically optimizes the number of bins and specific codecs used at each stage of the compression. If the user chooses to, however, they can manually control each of these choices.

For a consistent domain-specific behavior, one can integrate any global or local multi-level thresholding approach via an input image mask. Regions can contain overlapping intensity ranges and are determined on a per-pixel basis, so one can apply advanced pipelines to AGCR easily. To expedite this functionality, we implemented AGCR to admit Python "template" that apply any strategy (ex: multi-level Otsu thresholding[30]). For complex images, pixel-perfect regions can inflate file sizes, thus we provide an approximate mode as well. Lossless compression is still fully achievable in this mode but input regions from multi-level thresholding are no longer represented with a pixel-perfect contour. This sacrifices some compression efficiency for a dramatically smaller storage of shapes. The automatic configuration of AGCR integrates the approximate technique following a Gaussian blur to simplify contours.

As a representation method, the motivation of our technique is most comparable to that of the Adaptive Particle Representation (APR)[6]. Cheeseman et. al. argue that for representing fluorescence microscopy, a uniform grid of pixels should be replaced by particles which store intensity, image structure, and local resolution. This approach frontloads image pre-processing decisions but enables faster visualization and processing time while inherently decreasing file sizes. APR represents a content-adaptive disjoint partition of the image domain through a quadtree (2D) or octree (3D) structure. At high resolutions and lossless representations, a particle is placed at each voxel. This proves to be costly at the interface of high- and low-resolution regions as is common between sample and background edge transitions, including holes within the structure. Because particles are implemented with an underlying quadtree-based partition, every particle is axis-aligned and inherently follows a grid-like structure that the representation claims to avoid. Thus, we propose an alternative approach that can take advantage of the improvements in resolution representation introduced by Cheeseman and colleagues. We commend the motivation and implementation of APR and offer an entirely novel partitioning logic that is better suited to irregular biological images and extendable beyond the biomedical domain. We demonstrate our space partitioning strategy as compared to a quad-tree derived structure in Figure 2.

Just as APR particles are proportional to the number of input pixels and entropy of the data, a geometric contour representation reflects multi-resolution regions with a graph $G(V, E)$ comprised of a set of vertices $V$ and implicit edges $E$ for each 4-connected component. These components envelop a region and offer the similar benefit of separating physical objects in the data via the representation schema. A vertex-based representation benefits from a multi-resolution approach that down-samples continuous regions and is more succinct in doing so than a grid-based technique. Thus, geometric contouring fulfills all the representation criteria (RC) that APR posits: we guarantee a user-controllable representation accuracy for both noisy and noise-free images, our method's computational cost is proportional to the number of voxels, and our method offers similar potential for image processing and visualization independent from the original full-pixel representation[6].

Our partitioning method enables multiple pre-processing strategies that can benefit compression of images with high dynamic range and/or a high resolution. Using geometric contours, we modify or encode regions of the original image individually. With such an approach, we can separate complex sample regions from a largely continuous background and apply compression separately. By grouping an image into $k$ bins, we first reduce the range of possible intensity values which guarantees a reduction of variance when $k > 1$. Furthermore, when histogram packing is applicable, we can decrease both the variance and mean value of a given pixel. We can improve coding efficiency in separated regions, because linear and nonlinear approximation error decrease with the variation of an image[31]. To apply entropy encoding, which is the ultimate step in most lossless compression schemes, the optimal code length for a given symbol is $-log_b P$ where $b$ is number of possible discrete values and $P$ is the probability of a value. Intrinsically, separated regions have smaller values of $b$ and smaller resulting optimal code lengths than the original image. Thus, AGCR applies binning with one of two strategies aiming to benefit lossless compression of high bit-depth images. The first enables a distinct image compression or universal compression codec to be applied to each subset of the image raster. Due to the reduction in variance, this method is more effective for higher values of *k*. Furthermore, this "binned" approach enables the individual retrieval of image regions from a compressed file. The second strategy stores the modified pixel values in-place and applies a compression codec on the resulting raster. This "in-place" technique is especially advantageous for lossless image compression codecs that take into spatial information into account. As a result, lower values of *k* are more beneficial, reducing the number of discontinuities at the interface 4-connected regions. Only the binned strategy enables the fast retrieval of individual regions and application of multiple codecs because it compresses bins independently of each other. In a near-lossless or lossy context, the performance benefit of the application is much simpler. Improvements in compression ratio are proportional to the amount of tolerated loss and the corresponding size of the region(s) of interest.

## AGCR "Plus"

Alongside lossless compression, our method of geometric contouring enables a new paradigm of domain configurable near-lossless compression. AGCR separates regions of an image and enables the processing and compression of each region independently. In this manner, we enable the user to consistently compress images with a state-of-the-art, domain specific strategy while maintaining up to lossless accuracy in regions of interest. This capability ensures maximum compression performance without sacrificing data integrity. General lossy to lossless compression sacrifices context dependent advantages for a consistent ease of use. User tolerance for loss is configurable globally, and effective at a pixel-level resolution. In clinical and research contexts where the amount of loss acceptable is not standardized across image modality, even domain-specific compression strategies are dangerous because not every image in a domain shares the same characteristics. Thus, it is necessary for a bioimaging codec such as AGCR to support customizable, region-based compression on a per-image basis in order to avoid unpredictable data loss in critical areas.

We implemented this concept of progressive user configurability in what we term *AGCR+*. Users can specify the amount of loss via JPEG2000 for up to 256 sets of pixels in an image. Here, to contextualize the limit of 256, it is useful to think that methods with foreground & background use only two such sets. Threshold regions do not have to be connected and can contain overlapping intensity values. Theoretically, any lossy codec could replace JPEG2000, and the number of possible threshold regions could be extended. Pixel sets can be used to separate and compress user-defined objects regardless of their intensity ranges. For most practical cases we have observed that compression is most effective with 2-3 threshold regions. In Figure 3, we demonstrate the functionality of AGCR+ with and without templates as compared to other lossy methods. Regions of interest defining the sample can be manually or computationally drawn and permit fully lossless reconstruction of an irregular sample region. Because geometric contouring is concave, the background can be compressed with maximum efficiency. In contrast, a rectangular region of interest scheme would waste critical space for lossless compression for any non-rectangular ROI. Furthermore, as illustrated in Figure 3(A), we support progressive near-lossless compression of regions such that near a sample, a lower tolerance for loss can be employed.

Performance benchmarks

We validate our method and first demonstrate its near-lossless and fully lossless compression performance on medical and fluorescence microcopy images. In this domain, AGCR can provide immense improvements (upwards of 85%) over lossless compression codecs. For images larger than 600x600 pixels, AGCR performs at worst 0.58% worse than the best alternative compression method (2.07% for those under 600x600 pixels). Thus, even with exhaustive testing of each codec, the potential benefit of using AGCR outweighs the cost. Furthermore, for some images, universal compression is better than image-specific compression as evident in Figure 4(G, H). AGCR integrates the best of both compression strategies to offer comparable or better performance than each alternative. We stress that this technique is not contingent on the codecs used, on the contrary, AGCR would benefit from future codecs in the same manner.

Typical single-channel medical images have a bit-depth of 10-16 bits and are less than 512x512 pixels in size[8]. Breast tomosynthesis and chest radiography are significantly larger at 2457x1890 and 2000x2500 respectively for a 2D slice[8]. In contrast, the attributes of fluorescence microcopy images are less bounded. The size of the image raster produced in this modality varies depending on the microscope type and configuration[32]. To address each image modality, we include a range of datasets, each with distinct characteristics (Supplementary Figure 1). We collected twenty representative slices or separate images (when applicable) from high bit-depth medical images and microscopy images. The set of medical images is comprised of 140 knee and brain MRIs [33,34], Chest X-rays[35,36], CT Colonography[37], and MG mammography images[37]. Microscopy images (totaling 220) include a mEGFP-tagged Nucleophosmin (AICS-57)[38], PEGASOS-cleared brain[39], KRAS cell morphology images[5], drosophila embryo[40], cultured neuron[40], pan-Expansion (pan-ExM) image[41], and additional images from *Chakraborty* and colleagues[39]. We highlight that the automatic configuration of AGCR is comparable to or better than the best method of lossless compression for a given modality.

In Figure 4(A-C) we include the difference in compression ratio between the automatic version of AGCR as compared to XZ, BZ2, and JPEG-LS. We notice significant improvement of AGCR over universal compression techniques (A, B), increasing in performance as Shannon entropy increases. This follows the primary benefit of the binning technique that AGCR employs, separating intensity ranges into bins and histogram packing to reduce entropy for universal compression. JPEG-LS (C), as an image-based codec, proved to be more complex when determining the key characteristics of an image that made AGCR more effective. Here, we report dynamic-range normalized contrast (as measured by standard deviation) to illustrate that when competing with JPEG-LS, AGCR performs best on images with high contrast. Similarly, this supports AGCR's design to compress distinct regions with disparate image characteristics using multiple codecs at once. In Supplementary Figure 2, we expand each of these comparisons against each codec with additional dataset characteristics as the x-axis. In Figure 4E, we display the mean compression ratios for default ACGR, JPEG-LS[42], BZIP2[11], and XZ[43]. In addition, we demonstrate the near-lossless capability of AGCR lossless foreground *plus* lossy background compression (AGCR+). As described, AGCR+ is compatible with a user-assignable thresholding strategy and loss configuration. Thus, we include AGCR+ runs here as an example: users should determine an appropriate pipeline for their dataset modality. As discussed, lossless compression codecs have varied performance due to their underlying architecture. We suggest that AGCR is advantageous over independent compression codecs on two accounts. First, it provides the exhaustive comparison and combination of multiple independent codec. Second, it reduces entropy between regions of an image to offer potentially drastic improvements. We observe these advantages in Figure 4(G, H) for medical and biological images. For medical images, we see that the relative compression ratios of each codec can vary drastically between each images but AGCR's performance is consistently comparable or better to the contenders. For biological images, differences in compressibility are more uniform and distinct. The degree of separation differs per dataset, but JPEG-LS and AGCR tend to provide the best performance. We include runtimes in Supplementary Table 1.

To better illustrate the benefit of improvements in lossless compression in the medical field, we investigate the mammography (MG) image modality for which we consistently demonstrate compression improvement. On average, the raw MG images that we tested against were 27.81 MB. With AGCR, we saw a 49.53% improvement over JPEG-LS in lossless compression, equating to a reduced size of 5.36 MB vs. 8.89 MB respectively. The Mammography Quality Standards Act's (MQSA) program in the United States requires the lossless storage of mammograms for a minimum of 5 years[44]. As of March 1, 2021, the MQSA reports 38,878,310 annual procedures with at least 4 images per procedure[45]. Thus, for the typical clinical MG image size of 400 MB[8], this equates to 62.205 petabytes of uncompressed storage for one year alone. Following a linear projection, AGCR could reduce this total to 11.41 petabytes, saving 50.795 PB of compounding storage each year.

In evaluating the performance of AGCR, we examine two critical aspects of our technique. First, apart from histogram packing per region, AGCR as a compression technique is contingent on the performance of the codecs used. Here we include JPEG-LS, BZIP2, and XZ because they are accessible and have shown promise over other lossless compression techniques[8,10]. Second, for AGCR to be effective, an image must contain distinct regions to merit the separation of regions. Thus, for small images, the advantage of separating and compressing isolated regions is negligible or non-existent. This is reflected in Figure 4C where the global benefits of image-based compression outperform the reduction in entropy introduced by separating regions with AGCR. For images with high variance (Figure 4D), AGCR dramatically outperforms JPEG-LS because continuous regions of these images have disparate characteristics. In summary, AGCR can be a valuable compression scheme for high-resolution images with significant region continuity and disparity. The easiest measures of the latter characteristic are image dynamic range (typically larger for higher bit-depths) and image contrast (standard deviation). We also note the Synthetic Brain dataset as a high-resolution outlier with a bit-depth of 255, where AGCR still demonstrates a maximum of 34% improvement over JPEG-LS. Thus, one should not only rely on a single measure.

We solve the issues of variable codec efficiency and region determination by providing an automatic mode of AGCR. Manually tuning AGCR can yield even greater improvements, and we recommend it for advanced users operating on only a single image modality. However, we recognize that most users would not configure the system manually, therefore in all comparative analyses we only used the automatic mode to reflect the real-world benefits of AGCR.

Parameter Optimization

In the automatic mode of AGCR we balance and test the efficacy of various parameters, codec combinations, and codec-independent compression strategies. Thus, without user input, our automatic configuration offers an accessible, near-optimal integration of AGCR that is specific to each image. We select parameters at runtime through parameter-specific testing. For k-level thresholding, we iteratively reduce the number of bins tested, until sufficient Gini-based coverage is exhibited by each bin. Next, we determine a Gaussian standard deviation to simplify image regions by optimizing the number of resulting 4-connected regions. Finally, after automatically trimming shapes by size relative to the image's dimensions, our choice of compression strategy is determined through exhaustive testing. Together, these steps address the limitations and advantages of a given image by effectively choosing threshold, contour, and compression strategies.

The automatic configuration of AGCR selects the compression strategy on a per-image basis to address the variability of each technique. We indicate the impact of using AGCR in conjunction with other standardized lossless compression methods and codec-independent strategies in Figure 5. The choice of compression type depends on the efficiency of image-based codecs for a given image or set of image regions. JPEG-LS, for example, relies on the residual of a predicted image to follow a geometric distribution[42]. In this regard, we are balancing the region-specific use of two compression aspects: 2D vs 1D and global vs local pattern recognition. Binning an image favors 1D and local pattern recognition by reducing entropy and reducing the set of input intensities to a smaller range of repeated values. In this scenario, a higher number of thresholding bins tends to result in better universal compression but can drastically increase the number of small contours and computation time for AGCR. For the 2D and global methods, AGCR is best constructed with fewer bins to reduce inter-regional discontinuities. We observe in Figure 5 that the in-place versions of XZ and BZIP-2 never outperform their binned counterparts. This is due to the lack of two-dimensional pattern recognition for universal techniques. Thus, we found the best performance of AGCR to occur with universal techniques in a 'mixed' strategy. This mixed configuration combines the best of both aspects, binning high entropy regions with XZ and/or BZIP-2 and leaving the rest for JPEG-LS. When AGCR outperforms the in-place JPEG-LS configuration ("LS" in Figure 5), it is with this multi-codec strategy. We note that in a mixed mode of AGCR, JPEG-LS is applied to a binned region by cropping and padding with zeros after packing intensity values (Figure 1C). Altogether, by analyzing the various modes of AGCR, we bolster its most potent advantage: the application of multiple codecs to different sub-regions within a single image.

Domain Extensibility

In Figure 6, we demonstrate that AGCR can be applied to a variety of image modalities and unique domains beyond bioimaging. AGCR performs better than XZ, BZ2, and LS for 79.0%, 88.5%, and 97.3% of non-biomedical images respectively. Notably, our method proves effective for datasets with natural, irregular, and depth-based images. We include 220 indoor Kinect depth images[46], outdoor disparity maps[47], stereo matching depth and disparity maps for autonomous driving[48], underwater range maps[49] and separated underwater stereo RGB channels[50], categorized natural images (RGB merged)[51], and nighttime satellite images of the Earth[52]. In Figure 6, we highlight a significant improvement in compression ratio for AGCR compared to JPEG-LS. Depth and disparity maps are up to 126.50% smaller with AGCR than JPEG-LS. For this modality, the competitors to our method are the non-image-based compression codecs XZ, and BZIP-2. In contrast, when images are more continuous such as in the MIT FiveK datasets[51], AGCR is 113.02% better than XZ and 108.39% better than BZIP-2, where JPEG-LS is the competitor (+11.48% improvement). Regardless of the method, however, for 16-bit raw camera photographs, our performance is consistently better across subject types including people, nature, manmade objects, and animals. In summary, these results demonstrate that AGCR can easily be extended past biomedical domains to offer competitive lossless compression.

# Discussion

AGCR is a novel technique that partitions arbitrary regions of an image using vertex-minimized geometric contours. We demonstrated that this approach can deliver a significant improvement in lossless image compression, especially for high-resolution and high-contrast medical images, depth-based images, and high bit-depth natural images. We highlighted potential applications to visualization, retrieval, and progressive near-lossless compression. Finally, we highlight the improvements of AGCR's pixel-perfect representation over the rectangular model of the only previous bioimage representation, APR.

The utility of AGCR from compression stems from its ability to leverage the strengths and weaknesses of disparate methods within a single image. Regions with high entropy, such as backgrounds, benefit from universal encoders such as BZIP2[11]. In contrast, signal-rich foreground regions have content that is better suited for a predictive image-specific method like JPEG-LS[42]. Normally, the use of a single method across the entire image implies that, invariably, some part of the image is compressed with a suboptimal method. AGCR removes this inefficiency and allows optimal codecs and compression strategies to address each sub-region. As such, AGCR's performance is highly coupled with the combined use of multiple codecs and strategies (in-place, cropped, and binned). This is evident in the performance of our "mixed" modality as shown in Figure 5.

To maximize the impact of AGCR in practice, we were careful to create an automatic mode that performs well. As with most compression tools, a precise configuration of AGCR's available parameters can result in superior compression performance. However, manual or exhaustive parameter tuning is rarely done in practice, with most users opting for default settings. Therefore, we built AGCR with a robust automated mode by default to optimally determine any parameters that are not specified. Here, we solve multiple problems such as the number of thresholding bins, histogram coverage, and the tradeoff between region boundary quality and contour complexity. We tune multiple such parameters based on the nature of each individual image. AGCR automatically resolves these choices from the compressed file and correctly decodes the image. This content-adaptive strategy ensures that the performance on each file has maximal quality with minimal user input.

In future work, we suggest that one could effectively apply AGCR to both visualization and retrieval of images in 2D and in higher dimensions. We already enable the extraction of region contours in the Wavefront OBJ format, opening avenues for immediate visualization in 3D rendering pipelines. Contour visualization from an archived file could also be beneficial as a preview for inspection and/or retrieval of an irregularly sample region.

The visualization benefit is especially pertinent for super-resolution images where file input and output operations are computationally intensive. To this effect, geometric partitioning enables an efficient strategy for the lossless retrieval of a specific range of intensity values from the raw image. With such an approach, the user does not have to decode the entire image to access an area of interest. Extracted images can be compressed externally, decoded, and the recombined with the original remaining regions. This facilitates smaller file sizes along with faster encoding and decoding while maintaining data integrity. In an archival context when no loss is tolerated, such a strategy enables the use of a high-quality lossless compression for the background of a large image. For such a case, AGCR's ability to separate sample from background enables fast decoding of the foreground using a less efficient lossless compression scheme.

With further development, our geometric contouring algorithm could also be extended to volumetric (3D), time-series (4D), and multi-color (5D) imaging to more broadly address storage and retrieval of large bioimages. To this regard, we envision the application of existing state-of-the-art lossless video codecs in the context of pixel-perfect or shape-reduced contours. We anticipate that the performance benefits would be even greater in those regimes because while two or three regions suffice in 2D images, our method could be easily modified to support any number of regions. For super-resolution 5D volumes, our method's flexibility in categorizing diverse regions would likely prove to be even more useful.

In conclusion, we introduce the AGC algorithm through its example representation, AGCR, as a highly configurable, multi-faceted approach for compression, retrieval, and visualization of image data. In this study, we primarily investigate the application of AGCR to biomedical image compression. To this effect, our implementation enables the integration of image-specific thresholding schemes and exact specification of compression parameters. We also introduce AGCR+ which enables lossy to lossless compression of independent, user-defined regions. Finally, for lossless compression, we report comparable or better performance of the automatic version of AGCR to existing state-of-the-art lossless solutions across multiple distinct datasets. Thus, with geometric contouring, we demonstrate a significant advantage over raster-based storage and representations. Furthermore, this technique can be easily modified to integrate future state-of-the-art compression methods. We hope that our method lays the foundations to better high bit-depth and biomedical image-specific compression, visualization, and transmission methods.

# Figures

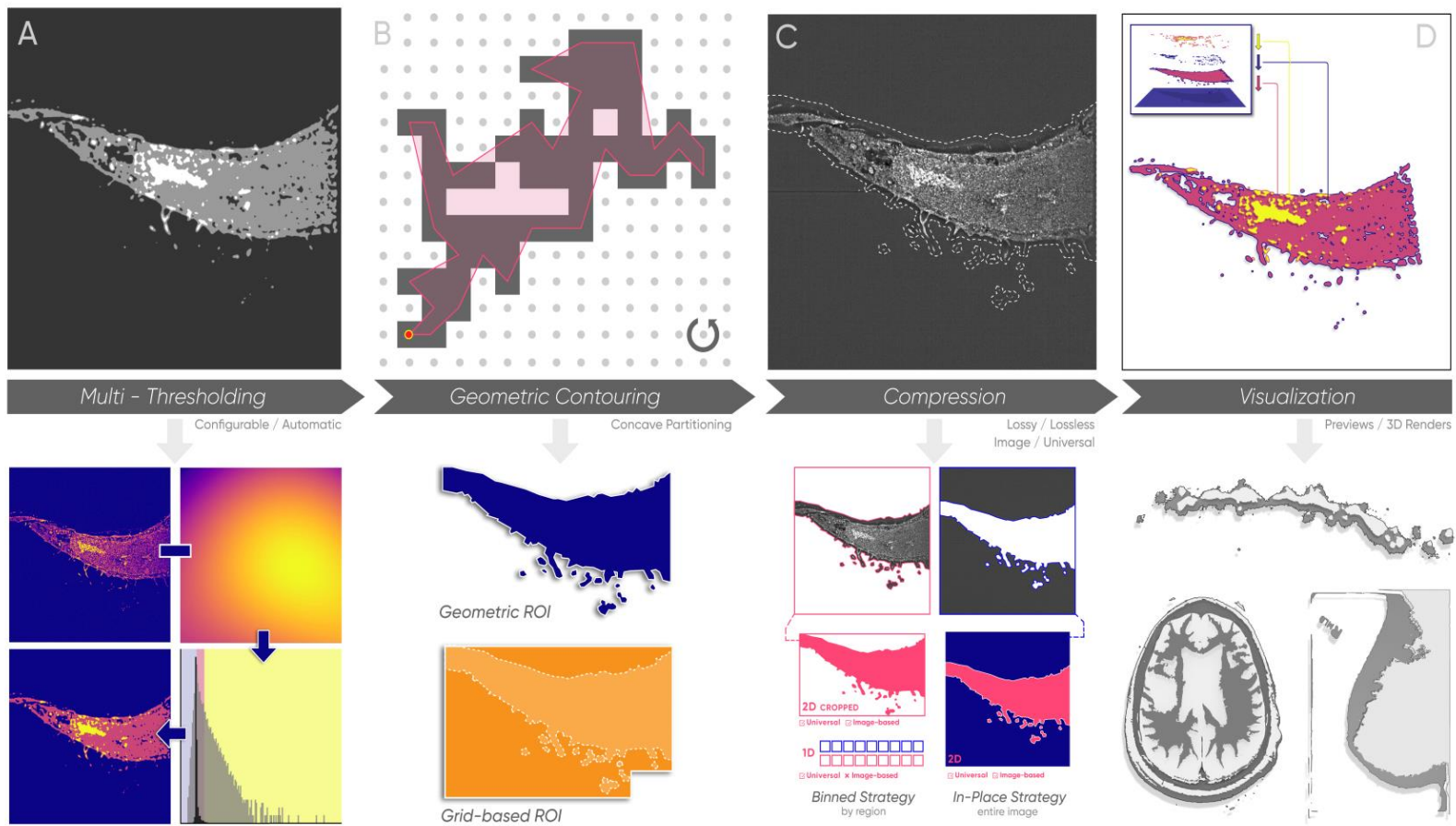

*Figure 1:* The Adaptive Geometric Contouring Representation (AGCR) achieves a per-pixel concave partitioning strategy for visualization, data elimination, data retrieval, and lossy to lossless region-based compression. We provide an automatic or fully configurable multi-thresholding approach to determine regions (A). Regions are wrapped with our novel geometric contouring algorithm: this stage can be estimated or pixel-perfect (B). AGCR supports lossy and lossless compression on a per-region basis (C). This is achieved through the binned and in-place application of existing image-based and universal compression codecs. We perform histogram packing on each region to minimize entropy and optimize compression. Finally, geometric contours can be visualized directly or extracted from a compressed file (D).

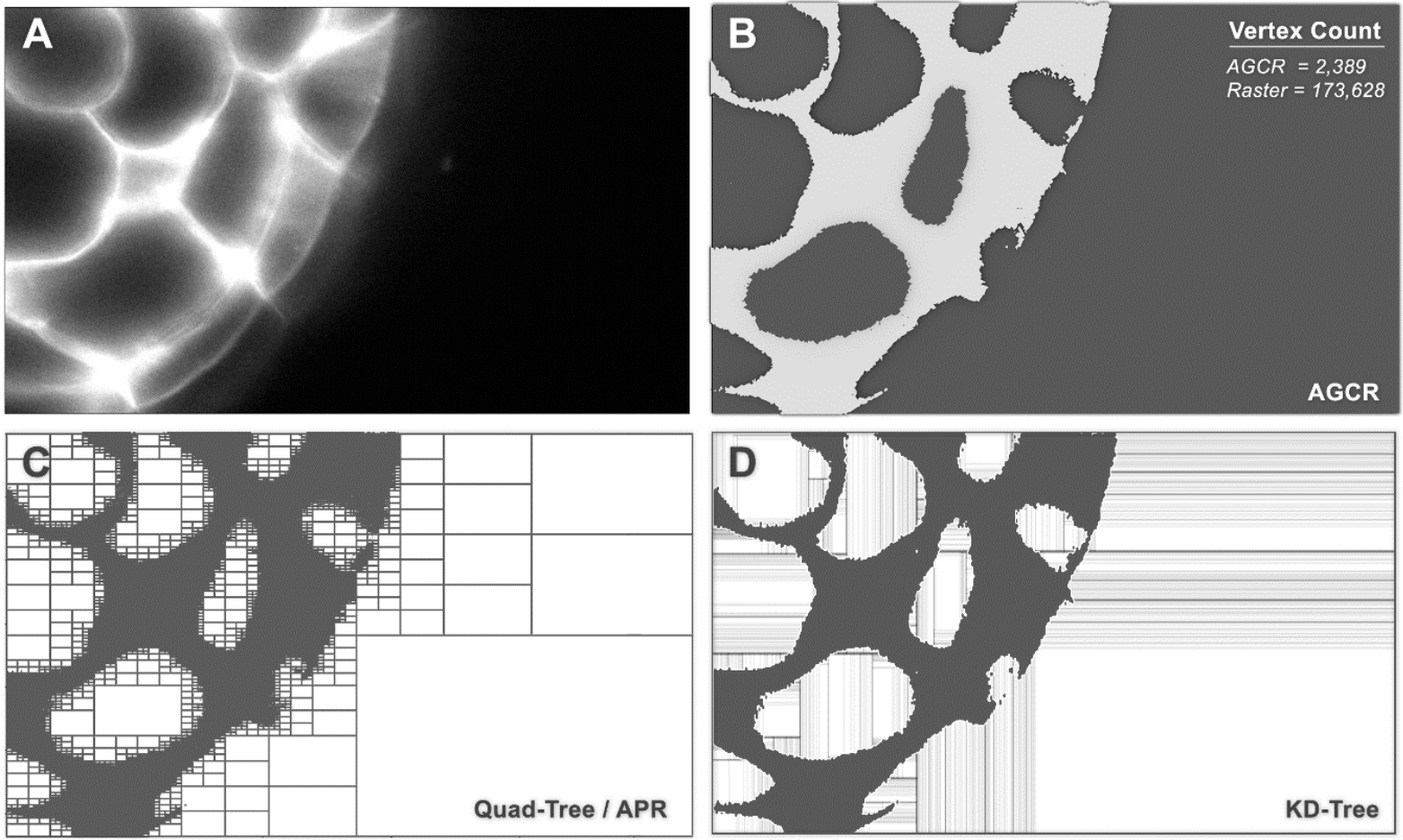

*Figure 2:* Biomedical images are characterized by irregular shapes that do not conform to grid-like space partitioning strategies. AGCR enables a more efficient representation of intricate, concave regions than achievable with Adaptive Particle Representation (APR). Depicted is a cropped raw embryo image (A) and a 3D visualization of the contour of AGCR at a given threshold value (B). We compare the partition complexity of AGCR (B) with a quad-tree as used in APR (C) and k-d tree (D) representation at a pixel-resolution contour. Furthermore, we demonstrate that AGCR results in a 98.62% reduction of vertices to represent the thresholded image compared with a raster approach.

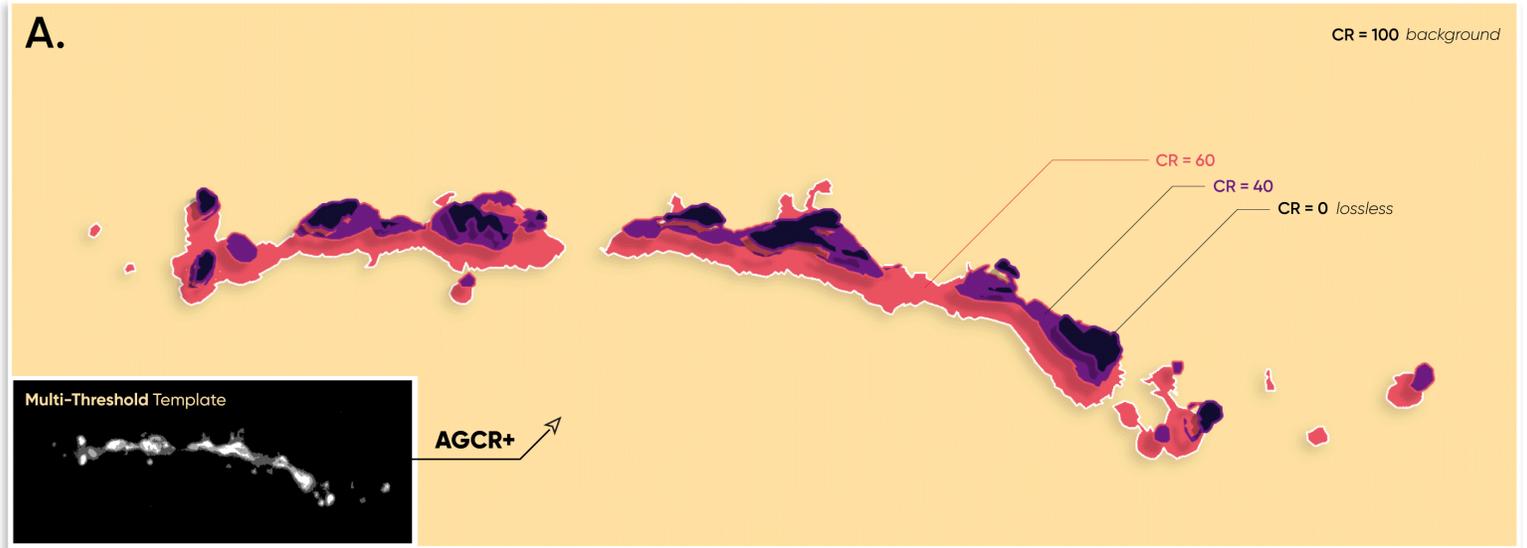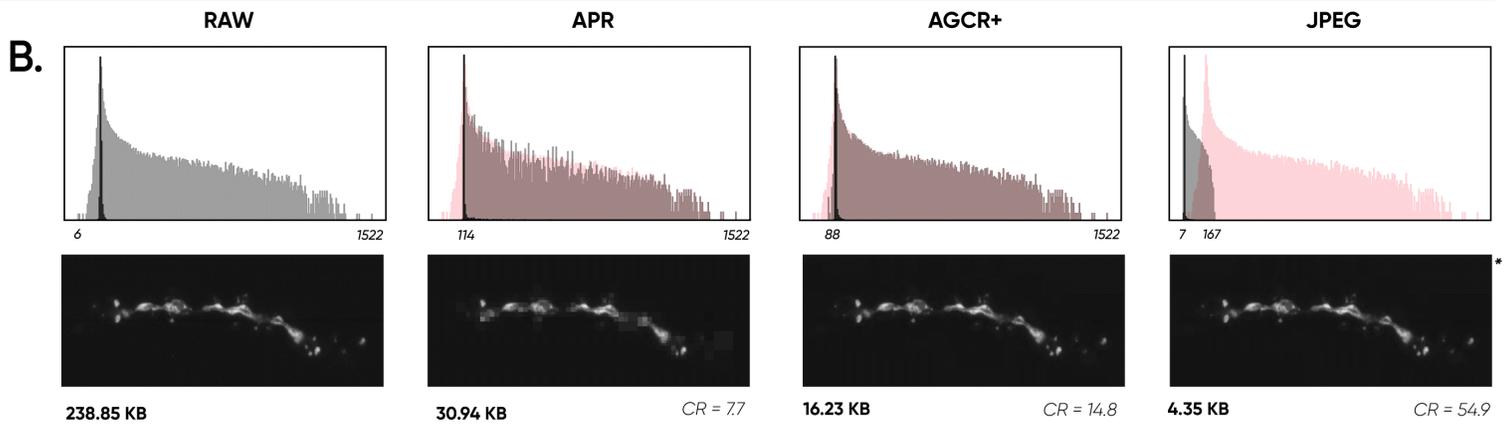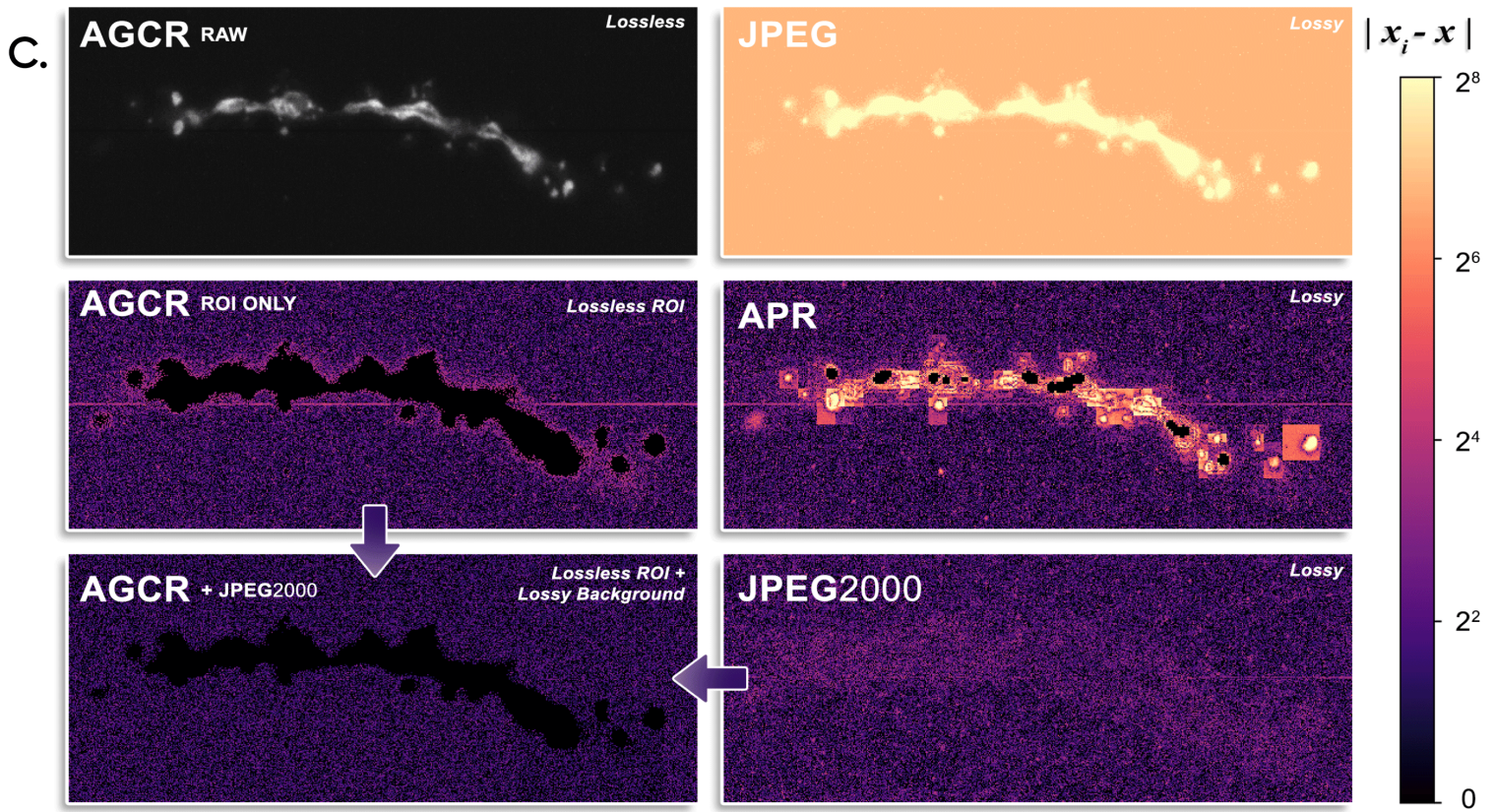

*Figure 3: AGCR+ enables context-dependent near-lossless compression that preserves sample integrity. By specifying a multi-threshold template, a user can employ our technique (A) with up to 256 threshold bins. With the example configuration, AGCR+ maintains the dynamic range and upper histogram integrity (B). APR also maintains the dynamic range but produces visual artifacts reflected in the histogram (y scale = $log_2$). Although visually similar when normalized, JPEG reduces the dynamic range and does not preserve intensity differences consistent with the original histogram\*. We demonstrate a similar effect in (C) with global binary thresholding. A user can compress the entire image lossless, or specify an ROI for lossless reconstruction. For the latter option, AGCR can store the background as the mean intensity value (ROI Only) or compress the background with a lossy codec (AGCR + JPEG2000). $Log_2$ absolute error is visualized between intensities of the raw image and each method (C). We observe significant loss for the sample region in JPEG, APR, and lossy JPEG2000 ("CR =30"). In general, APR preserves the regions with highest intensity but does not preserve the sample as well as JPEG2000.*

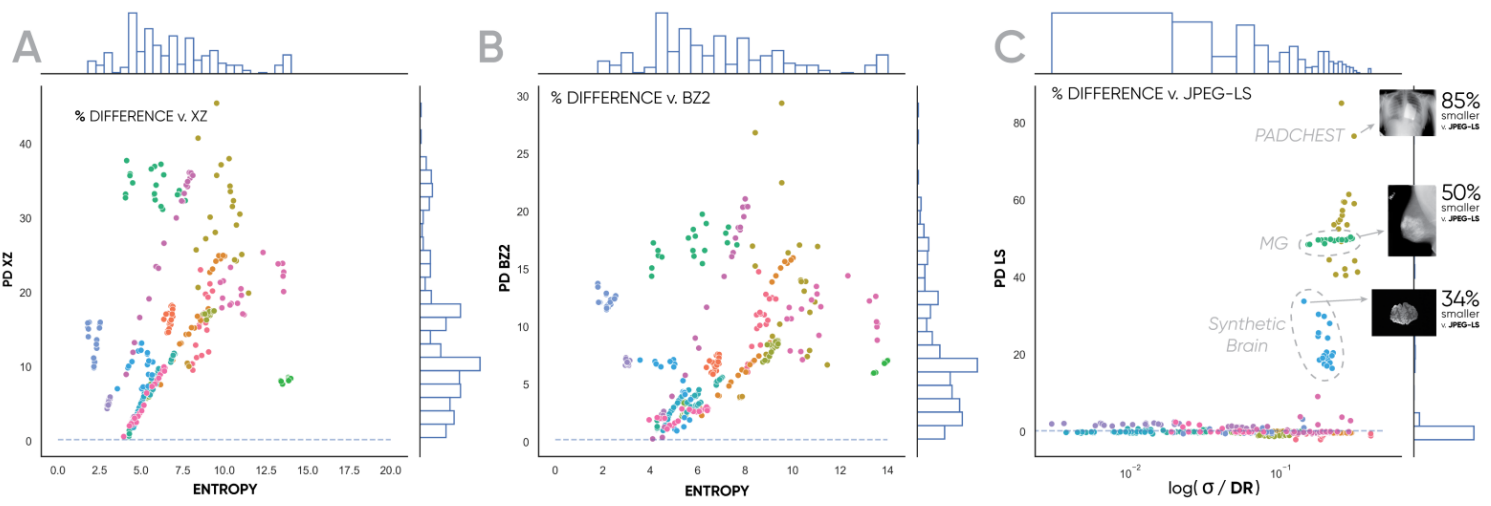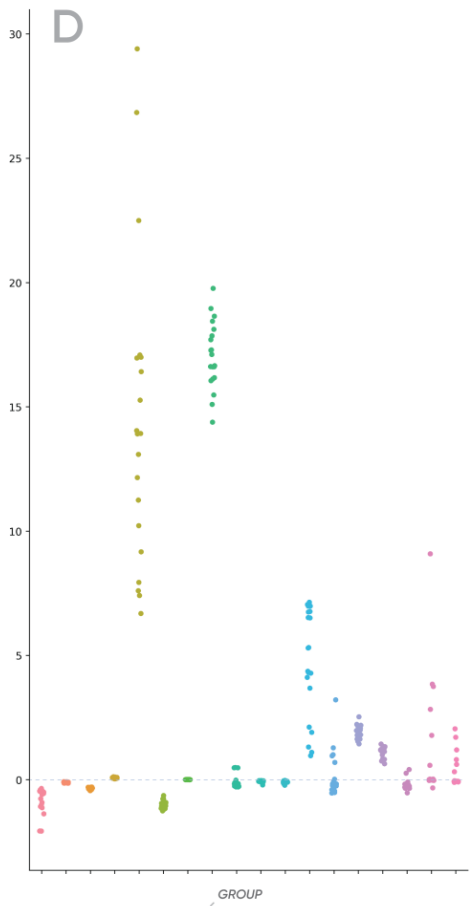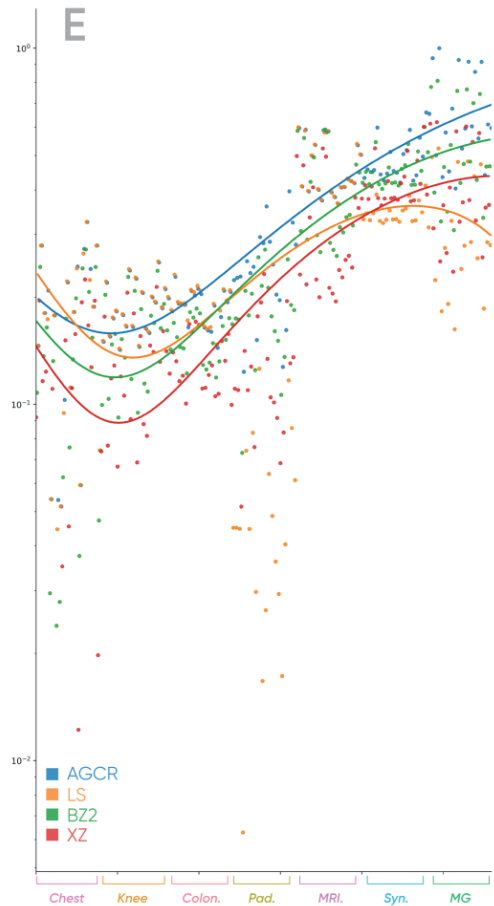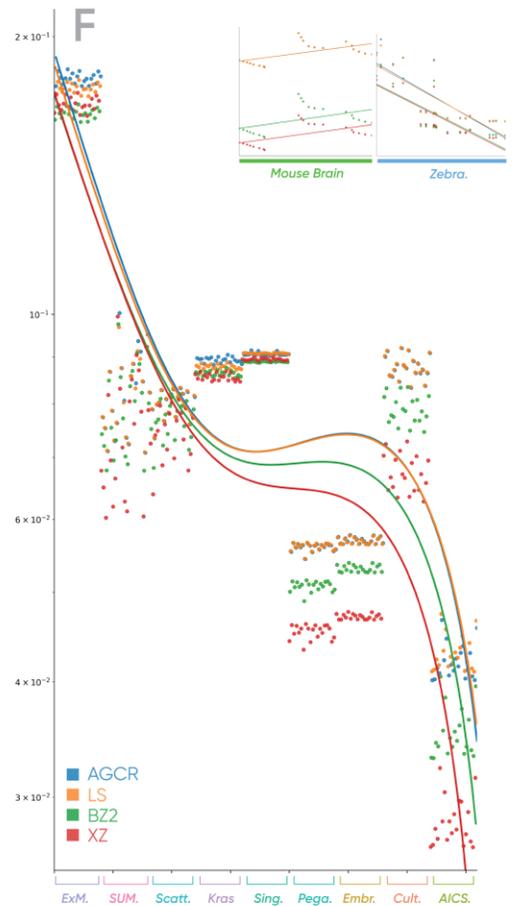

***Figure 4:** Displayed are the percent differences (PD) in compression ratio (CR) of the automatic mode of AGCR compared to XZ (A), BZIP2 (B), and JPEG-LS (C). For JPEG-LS, large images with higher contrast (PadChest, MG, and Synthetic Brain) perform 20-80% better with AGCR (C). We illustrate percent difference compared to the best codec on a per-image basis in (E). We display the relative compression ratio between lossless codecs for medical (E) and biological images (F). In (E), samples are sorted by AGCR's compression ratio, and in (F) samples are sorted by standard deviation. Finally, in (G) we display the average compression ratio for each dataset. On average, our AGCR performs +11.34% better than the best of alternative methods per medical image and matches the performance of the best on biological images, improving by +0.33%.*

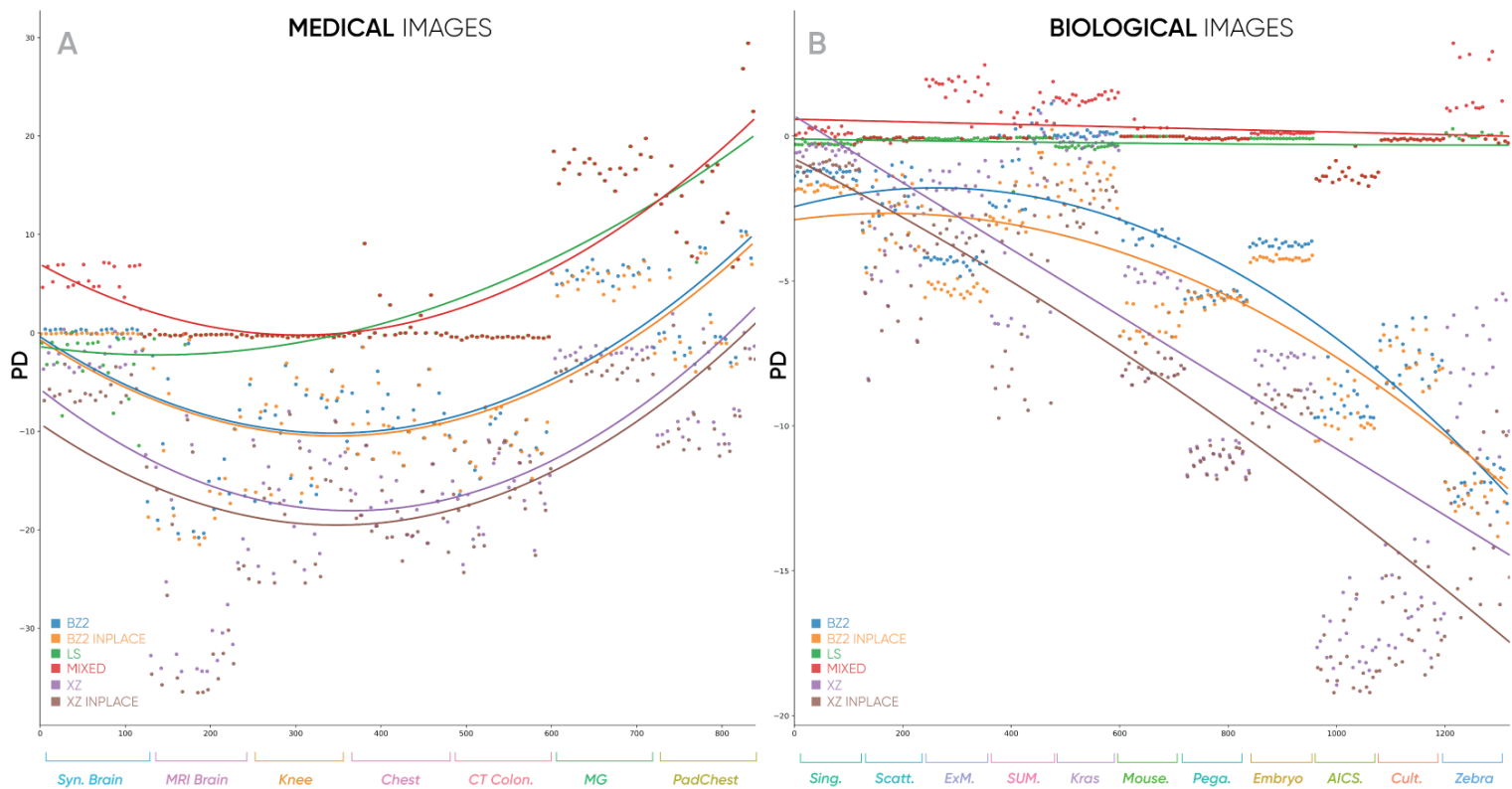

*Figure 5: AGCR operates best when applying multiple codecs to a single image.* We display different modes of AGCR, integrating in-place and binned strategies with various codecs. Medical images (A) are sorted by standard deviation, and biological images (B) are sorted by Gini index. Percent difference of each mode is reported as compared with the best of BZ2, JPEG-LS, and XZ on their own. Here we see the JPEG-LS and "Mixed" modes of AGCR perform the best across datasets.

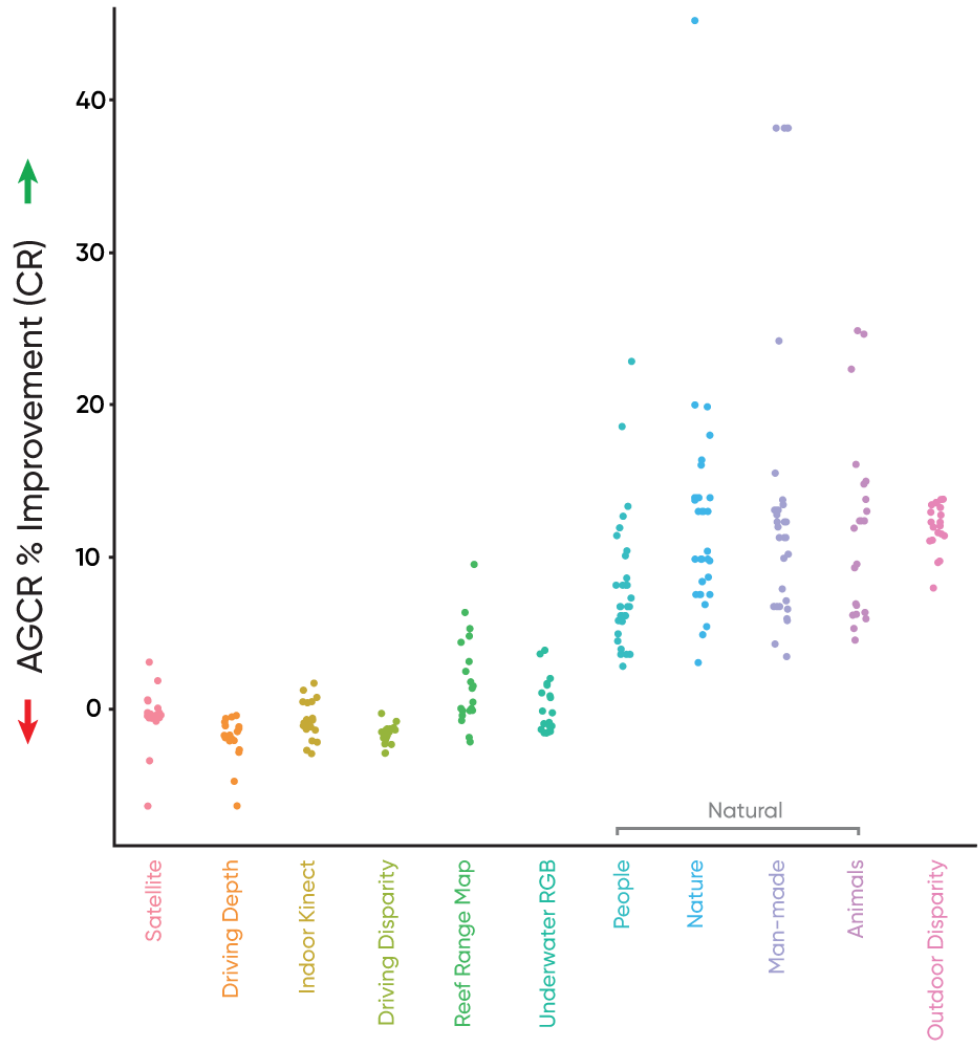

*Fig. 6: AGCR is effective for images beyond the bioimaging domain.* We demonstrate significant performance improvement in lossless compression for depth-based, natural, and additional images. On average, AGCR demonstrates a 57.3%, 58.0%, and 37.9% improvement in compression ratio compared to XZ, BZIP-2, and JPEG-LS across images. We plot relative performance against each codec in Fig. S6.

# Methods

AGC Theory and Algorithms

We have constructed a novel geometric compression algorithm, AGC, that produces a vertex-minimized contour, separating regions at a pixel level. We implemented the software in C++ to process 16-bit image rasters for lossless compression and visualization. The process is highly configurable with the following core components: image analysis, geometric contouring, contour optimization and reduction, region optimization, and then final encoding.

In the first step, we read the image raster from the input file format and validate the integrity of the file and its metadata. An initial image analysis is then performed, followed by *k*-level thresholding to determine region intensity ranges prior to geometric contouring. Our implementation uses LibTIFF[53] to read and process input images of the TIFF file format. For this purpose, we converted input files of alternative formats to 16-bit TIFF files prior to execution. Beginning the analysis step, we divide the image into range proportional bins and calculate the inverse Gini coefficient 1 – *g* to establish a minimum probability when constructing thresholding bins. We employ this strategy to measure the sparsity of the image, where *g=0* indicates that all pixels are of the same intensity and *g=1* indicates a maximal inequality of values. The minimum probability enforces a ceiling for *k* informed by the data and ensures that an appropriate number of values are present in each intensity range. Following this calculation, we perform histogram packing with a reversible transform that eliminates missing intensity values to construct a continuous histogram beginning at zero and ending at the modified histogram frequency N' - 1. We then recursively subdivide the image into bins of a minimum range of 2 bits. This range is parameterized and can be modified to construct bins within a given bit-depth. Subdivided bins are merged to achieve a minimum probability *p* for a given bin of frequency *f*, where *p = f / N'*. As an exception, if an input floor value is supplied for the image, all bins with a minimum value less than the floor value are merged. Prior to this step, the input floor value must be transformed according to the histogram packing. In the absence of a thresholding template, AGCR will apply Otsu's threshold method[30] for images with a low dynamic range or high Gini index to determine this floor value. If a threshold template is specified, only histogram packing and the accompanying transform are performed. As a result, for this step, we inform geometric contouring of regions based on their intensity ranges in accordance to the *k* calculated or provided threshold bins.

For the automatic configuration of AGCR, we approximate contours of intensity ranges by introducing a gaussian blur prior to region determination. This Gaussian filter acts as a built-in template that favors simplified regions. To reiterate, this step could be replaced with any user-generated template and strategy. For our Gaussian process, we optimize two parameters, the number of thresholding bins to choose and the standard deviation of the gaussian blur. For the former, we optimize the number of bins based on histogram coverage. Coverage for bin $b_k$ is defined as (1 – *g*)\*range($b_k$). If the probability of a bin $f_k / N'$ is less than its coverage and 1-*g*, then we reduce the number of bins and repeat this process. For the latter parameter, we search for the standard deviation value that minimizes the number of output regions. This is done using the number of bins previously determined to create and count 4-connected regions. While the number of regions are greater than 10*$k$, we increase the standard deviation by five. These parameters were chosen empirically but could be refined to smaller step-sizes for a more exhaustive automatic parameter determination at the cost of run-time.

In the second step, geometric contouring, we first cache the transformation between each raw intensity value and its index from *0* to *k-1* to improve computational efficiency. We term this value a "tolerance index," representing the index into a histogram bin containing the range of intensities that have been histogram packed and normalized based on the bin size. When a template image is provided, intensity values are decoupled from their respective tolerance index, so we do not need to cache or store index to value transformations. Next, we iterate through the pixels of the image raster, performing a depth-first search on Von Neumann, 4-connected neighborhoods with the same tolerance index. We check if a given position has been visited by 2D indexing into a 1D array of boolean values which offers faster performance when compared to an unordered set of vertices. Connected components are then wrapped with our concave sweeping algorithm which we term adaptive geometric contouring. We define the objectives of our method as follows:

   a. Minimize the total number of vertices that define a shape.
   b. Ensure shapes do not intersect.
   c. Maintain the regularity of a shape as much as possible.

These requirements are motivated for lossless compression and visualization of shape contours. Objectives (a) and (b) prioritize lossless shape compression, and (c) loosely defines a requirement for geometric contours generate shapes that can be triangulated aesthetically. Adaptive geometric contouring thus wraps regions counter-clockwise, minimizing the number of vertices while remaining within the bounds of the 4-connected region. This step is expanded in Algorithm 1.

In the third step, contour optimization and reduction, we iterate over regions and perform a second stage of non-destructive vertex optimization then, if specified, we destructively reduce the number of vertices in each shape to aid in shape compression. This first stage visits each vertex to find the longest possible edge that does not include vertices out of the region. This process is described in Algorithm 2. The second stage removes vertices based on a parameter that specifies the maximum number of vertices to keep per 100 pixels. Prior to vertex removal, we remove shapes with a number of pixels less than a minimum size *m = max(32, width\*length\*0.4e-5)* which addresses small and super-resolution images.

In the fourth step, region optimization, we organize and non-destructively optimize the size of the data that stores geometric regions prior to encoding. First, because regions of a given tolerance index can envelop smaller regions of a different index, we sort regions by their area. Larger regions are decoded first so that smaller, enveloped regions can be written on top of the resulting raster during the decoding process. We perform one such decoding step after first removing all regions of tolerance $t = 0$ and storing them temporarily in a list. We compare the decoded tolerance raster with the original tolerance raster and then push back any removed regions that cause an inconsistency. Finally, we resort the regions in descending order by area. This process results in a reduction of total vertices $V$ in the final graph roughly proportional to $V / k$. Because most raw medical and microscopy images contain a black background that ramps in intensity value at areas of the image closer to the sample, this technique writes shapes of higher intensity over a largely sparse background corresponding the tolerance index $t = 0$. Thus, we can usually assume an initial shape that covers the entire image with tolerance $t = 0$ and only store the outlier that represent enveloped shapes of this same intensity range. Such an approach can be adapted for images of a different modality and applied to any tolerance index when desired. When maximum compression is desired, we repeat this shape removal process for each tolerance range in the 'slowest' run configuration. Following these optimizations, we prepare the remaining shapes for encoding with multiple separated components. We store lists of region sizes, region tolerance indices, x positions, and y positions each separately. We calculate the bounding box for each region and store this global value along with local offsets for each vertex of the shape relative to the bottom-left corner of the bounding box. Additionally, we store the position of the corner relative to the previously processed shape. Because values are all unsigned, we bit shift these values to set the least significant bit to 1 for a negative difference and 0 for a positive difference between current and previous x and y positions. During this process, duplicate shapes are removed with a temporary shape dictionary. If a match is found between two shapes based on local offsets from the bottom-left bounding box corner, we store an index reference to the first shape occurrence. This is indicated by a region size of zero in the size list which indicates to retrieve the next index of the shape occurrence and copy its data. Altogether we optimize region size by minimizing stored values while maintaining original ordering of the data for lossless decoding. Similar components of the resulting output are grouped and compressed with XZ-Utils[43] or BZIP-2[11]. In a faster configuration, XZ-Utils is used by default, otherwise each is tested per component. Finally, we compare this output with a JPEG-LS compressed tolerance raster and use the smaller of the two schemes when –NOVIS is passed as an argument.

In the last step, we perform final encoding of the modified intensity values with one of two methods. The first method stores all intensity values in their original raster positions after histogram packing and bin normalization. We then employ lossless image compression or in-place universal compression to store the in-place intensities. The second method maintains the binned intensity values sorted in raster order within separated bins prior to universal compression. By default, we employ BZIP-2 to compress each bin, but a slower configuration will test each codec. For JPEG-LS to function in the binned mode, we "crop" the region and pad outside pixels with zero (see Figure 1). Finally, we store the histogram packing transform and offsets of each tolerance bin for decoding the original intensity values. When templates are provided or for AGCR+, we employ JPEG2000[54] for any lossy regions. Users can specify loss via the compression ratio per bin.

Decoding is a simpler process compared to encoding. We first reconstruct the tolerance raster $T$ including every region and its tolerance index by reversing the original transformations. After vertices are in global space, we iterate through each region, writing its vertices and edges to *T*. Then we iterate through the pixels contained in the bounding box of each region and check if they are also contained within the concave polygon using a modified version of the algorithm described by Finley[55]. For many regions across a super-resolution image, this step can be time consuming. Thus, we implemented a multithreaded option that segments regions based on the number of threads available and performs a threaded raster scan on each segment. It is important to note that in step 4 of compression, this threaded shape filling can be performed up to *k+1* times, accounting for each bin and a validation check. Following construction of *T*, we decode the intensity values, reverse the histogram packing, and then fill in the tolerance raster with the reconstructed values.

For visualization, the stored tolerance raster can be retrieved to view each region and its tolerance value. Alternatively, we can write the shape graph to a Wavefront OBJ file that can be imported and displayed in most 3D software. We include an example visualization in Supplementary Figure 5.

For lossless compression with AGCR, choice of effective parameters depends on two factors: the compression scheme used, and the characteristics of a given image. We first discuss the advantages and limitations of each compression strategy. We perform lossless encoding of the processed image with JPEG-LS[33], XZ-Utils[34], BZIP2[5], and JPEG-2000[37] at multiple configurable stages, but the application of each codec is realized through a distinct strategy of application. The in-place strategy primarily implements AGCR to selectively provide histogram packing across regions of an image. The locations of each region are implicitly stored, but only one codec can be employed to compress the image and the entire image must be decompressed to access a region of the image. The binned strategy stores regions in exact or overlapping intensity-ranges. In this scenario, multiple compression codecs can be used on the same image and individual regions can be retrieved independently of each other. However, the location of regions in the base image must be stored explicitly, favoring less geometric contours. The second factor, image modality, has a tremendous impact on the coding efficiency of AGCR and choice of compression scheme. Image sparsity and dynamic range affect the performance of histogram packing[38] and frequently serve as limiting factor for lossless image compression of medical images when compared to natural images[4]. Second, we note that in addition to histogram packing, AGCR depends on the complexity or "texture" of the image. Distinct and continuous regions are favorable for reducing entropy across each intensity range and applying different codecs. Isolated illuminated subjects are advantageous for the same reason due to a higher contrast, which limits geometric complexity. Finally, region-based histogram packing is especially effective for noisy backgrounds and, when properly segmented, this feature can vastly improve the performance of AGCR. Each of these image characteristics are encouraged by images with a high bit-depth and/or high resolution, an observation that is demonstrated in our results.

For AGCR to offer competitive compression ratios we needed to tackle the storage of geometric contours. On a surface level, geometric contouring minimizes the number of vertices needed to represent a shape. However, a pixel-perfect wrapping of high-resolution intensity ranges can create an immense amount of interweaved, complex concave shapes. These many regions are required to provide an exact segmentation for reducing entropy and providing regions specific to exact intensity ranges. Thus, the crux of AGCR's optimization problem involves the balance of high region complexity for optimal image compression and low region complexity for shape compression. We optimize storage on a per-shape level by various techniques, including the recording vertex offsets and a shape dictionary to help alleviate this concern (Methods). Still, the sheer volume of and complexity of shapes in an image with high variance necessitates a more advanced compromise. We introduce the concept of shape reduction following the construction and optimization of geometric contours to serve as approximate bounding contours with drastically less vertices. Furthermore, to help alleviate shape complexity, the automatic configuration of AGCR begins with a gaussian blur to simplify the contours of each intensity range. Both stages, however, lead to the overlap of these ranges and reduce the downstream efficiency of the lossless compression codecs chosen. A non-destructive alternative to increase shape efficiency is simply to reduce the number of bins at the multi-level thresholding stage. Thus, without shape reduction, we found binary binning to perform the best in practice. Furthermore, inter-region discontinuities are minimized with fewer bins, favoring the 2D pattern recognition of JPEG-LS. Choices of bins for the automatic and exact mode of AGCR are displayed in Supplementary Figure 4. "Exact" here refers to AGCR with no shape reduction or Gaussian template. Our automatic configuration weighs these risks and optimizes bin counts, multi-level thresholding, gaussian standard deviation, and shape reduction at run-time to predict the most efficient combination of parameters for exhaustive testing of compression codecs downstream.

For lossless compression with AGCR, choice of effective parameters depends on two factors: the compression scheme used, and the characteristics of a given image. We perform lossless encoding of the AGCR processed image with JPEG-LS[42], XZ-Utils[43], BZIP2[11], and JPEG-2000[54] at multiple configurable stages. The in-place strategy primarily implements AGCR to selectively provide histogram packing across regions of an image. The locations of each region are implicitly stored, but only one codec can be employed to compress the image and the entire image must be decompressed to access a region of the image. The binned strategy stores regions in exact or overlapping intensity-ranges. In this scenario, multiple compression codecs can be used on the same image and individual regions can be retrieved independently of each other. However, the location of regions in the base image must be stored explicitly, favoring less geometric contours. The second factor, image modality, has a tremendous impact on the coding efficiency of AGCR and choice of compression scheme. Image sparsity and dynamic range affect the performance of histogram packing[56] and frequently serve as limiting factor for lossless image compression of medical images when compared to natural images[10]. In addition to histogram packing, AGCR depends on the complexity or "texture" of the image. Distinct and continuous regions are favorable for reducing entropy across each intensity range and applying different codecs. Isolated illuminated subjects are advantageous for the same reason, and due to a higher contrast which limits geometric complexity. Finally, region-based histogram packing is especially effective for noisy backgrounds and, when properly segmented, this feature can vastly improve the performance of AGCR. Each of these image characteristics are encouraged by images with a high bit-depth and/or high resolution, an observation that is demonstrated in our results.

For AGCR to offer competitive compression ratios we needed to tackle the storage of geometric contours. On a surface level, geometric contouring minimizes the number of vertices needed to represent a shape. However, a pixel-perfect wrapping of high-resolution intensity ranges can create an immense amount of interweaved, complex concave shapes. These many regions are required to provide an exact segmentation for reducing entropy and providing regions specific to exact intensity ranges. Thus, the crux of AGCR's optimization problem involves the balance of high region complexity for optimal image compression and low region complexity for shape compression. We optimize storage on a per-shape level by various techniques, including the recording vertex offsets and a shape dictionary to help alleviate this concern (Methods). Still, the sheer volume of and complexity of shapes in an image with high variance necessitates a more advanced compromise. We introduce the concept of shape reduction following the construction and optimization of geometric contours to serve as approximate bounding contours with drastically less vertices. Furthermore, to help alleviate shape complexity, the automatic configuration of AGCR begins with a gaussian blur to simplify the contours of each intensity range. Both stages, however, lead to the overlap of these ranges and reduce the downstream efficiency of the lossless compression codecs chosen. A non-destructive factor to increase shape efficiency is simply to reduce the number of bins at the multi-level thresholding stage. Altogether, our automatic configuration optimizes bin counts, multi-level thresholding, gaussian standard deviation, and shape reduction at run-time to predict the most efficient combination of parameters for exhaustive testing of compression codecs downstream.

# Acknowledgements

We thank the support and funding of the Lyda Hill Department of Bioinformatics at UT Southwestern Medical Center. Kevin Dean and Reto Fiolka were instrumental in providing images and guiding dataset retrieval throughout the development and presentation of our method. Microscopy images provided by multiple collaborators were also integral to the implementation and analysis of our technique. We thank Meghan Driscoll for supplying the "Kras" dataset; the Bezprozvanny Lab (UT Southwestern Department of Physiology) for contributing the "Scattered Neurons" dataset; Ons M'Saad and the lab of Joerg Bewersdorf for providing the pan-Expansion Microscopy dataset; the lab of Ilya Bezprozvanny for the dataset labeled "Scattered Neurons"; and Wen Mai Wong, Cara Nielson, and Julian Meeks for providing the "Mouse Brain" dataset and additional microscopy images during development.

# Author information

## Contributions
The project was conceived by M.C. and K.V, algorithms and analysis by K.V., software implementation by K.V, additional theory & guidance by M.C..

# Ethics declarations

## Competing Interests
Two provisional patent applications have been filed in the U.S. Patent and Trademark Office regarding the methods detailed in this manuscript titled: SYSTEMS AND METHODS FOR CONTOURING (No. 63/350,674) and SYSTEMS AND METHODS FOR IMAGE AND VIDEO COMPRESSION (No. 63/329,826) with inventors Murat Can Çobanoğlu and Kevin Christopher VanHorn, assigned to The Board of Regents of The University of Texas System.

# Supplementary Tables and Figures

| Dataset | AGCR (sec) | AGCR+ (sec) |
|---|---|---|
| AICS-57 | 1.2 | 0.91 |
| CT Colonography | 1.76 | 1.71 |
| Chest | 46.80 | 45.62 |
| Cultured Neuron | 16.94 | 12.72 |
| Embryo | 11.10 | 8.17 |
| Knee | 0.99 | 0.73 |
| Kras | 1.79 | 1.17 |
| MG | 132.47 | 127.17 |
| MRI Brain | 3.62 | 2.62 |
| Single Neuron | 6.62 | 5.93 |
| PadChest | 60.11 | 54.70 |
| Pegasos Matrix | 45.63 | 42.42 |
| Sum 159 | 22.69 | 18.58 |
| Synthetic Brain | 27.59 | 21.86 |
| Mouse Brain | 335.07 | 349.61 |
| Zebrafish | 51.15 | 41.15 |
| Pan-ExM | 49.51 | 39.66 |
| Scattered Neurons | 85.25 | 80.48 |

**Supplementary Table 1:** *Mean runtime (in seconds) for the default "automatic" configuration of AGCR. Both methods were run with the with shape compression over 32 threads and the '--slowest' exhaustive compression setting.*

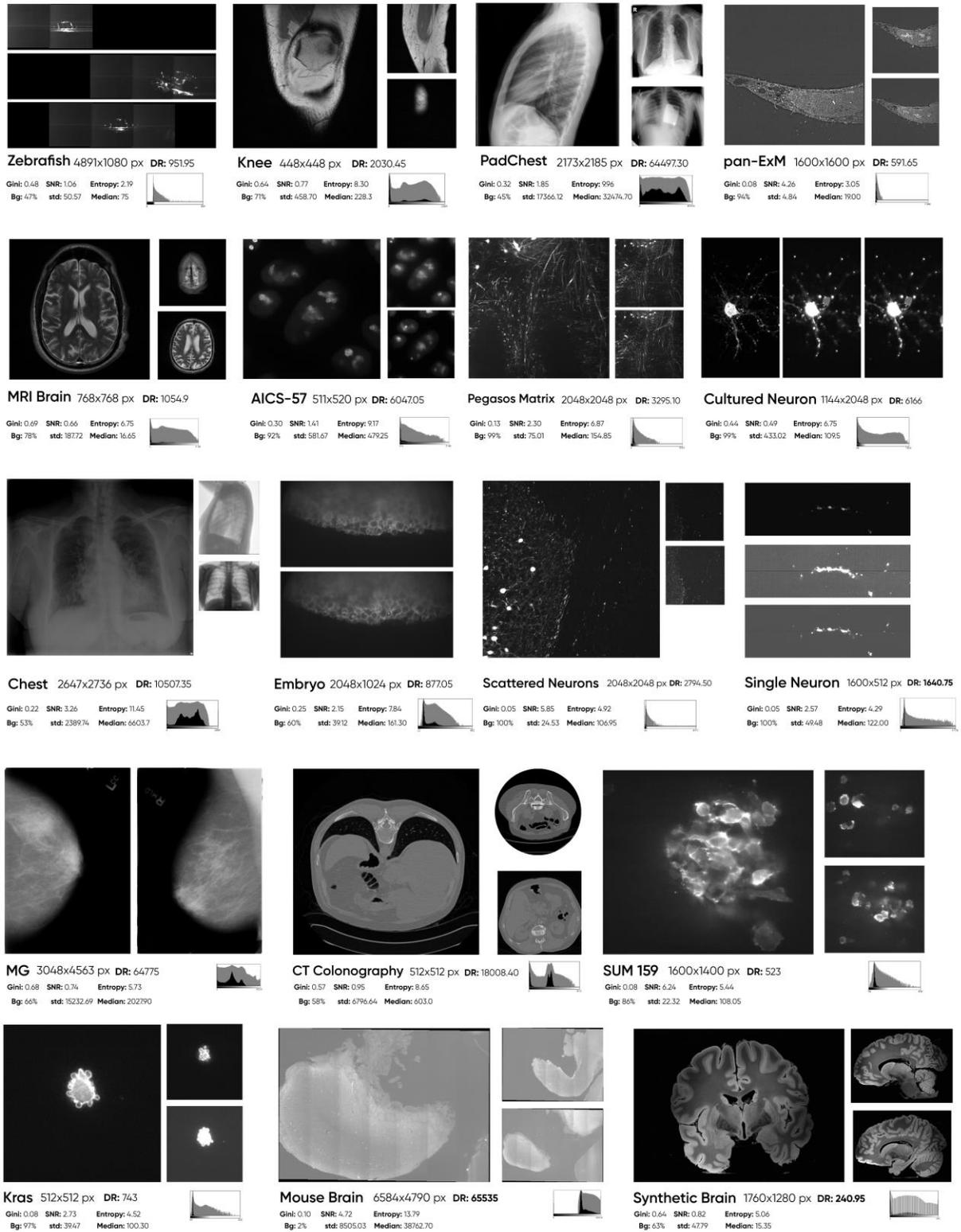

**Supplementary Figure 1:** We include a variety of biological and medical datasets to evaluate the performance of our technique. Depicted is the Gini index, signal-to-noise ratio (SNR),

Shannon Entropy, percent background using Otsu thresholding (Bg), standard deviation (std), and the median value of each image.

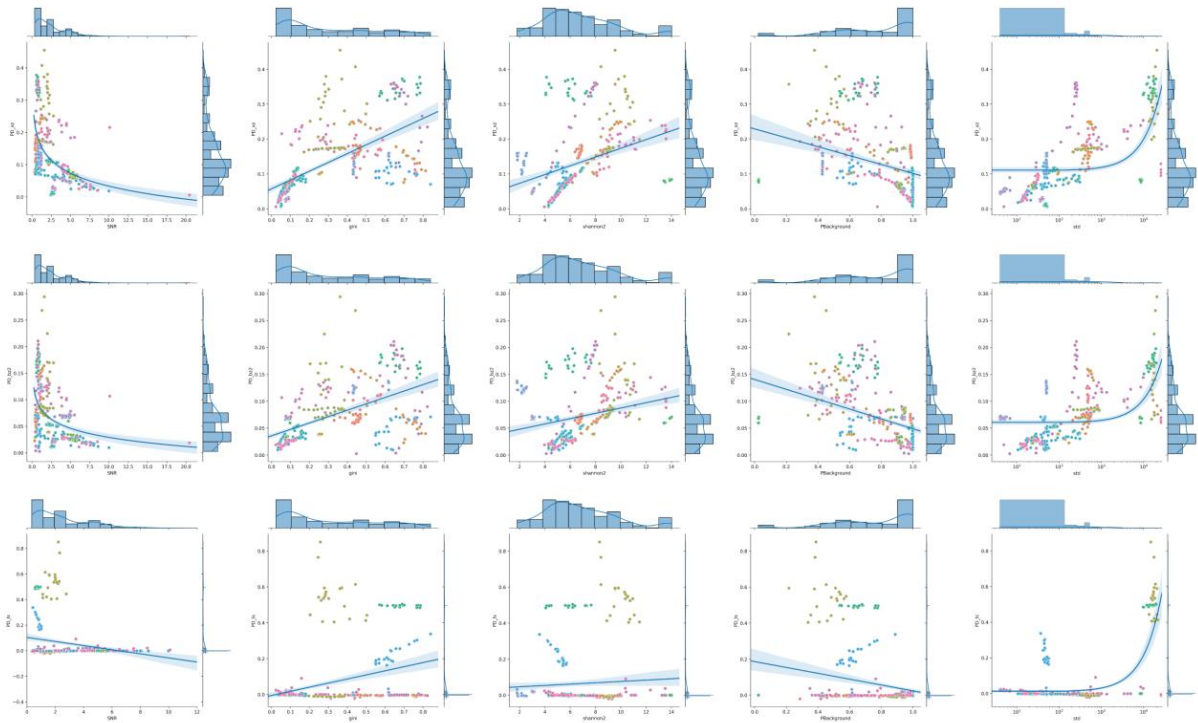

**Supplementary Figure 2:** *The automatic lossless version of AGCR is contingent on multiple image characters. Here we display the percent difference of AGCR against XZ, BZIP2, and JPEG-LS as measured by SNR, Gini index, Shannon entropy, percent background, and standard deviation.*

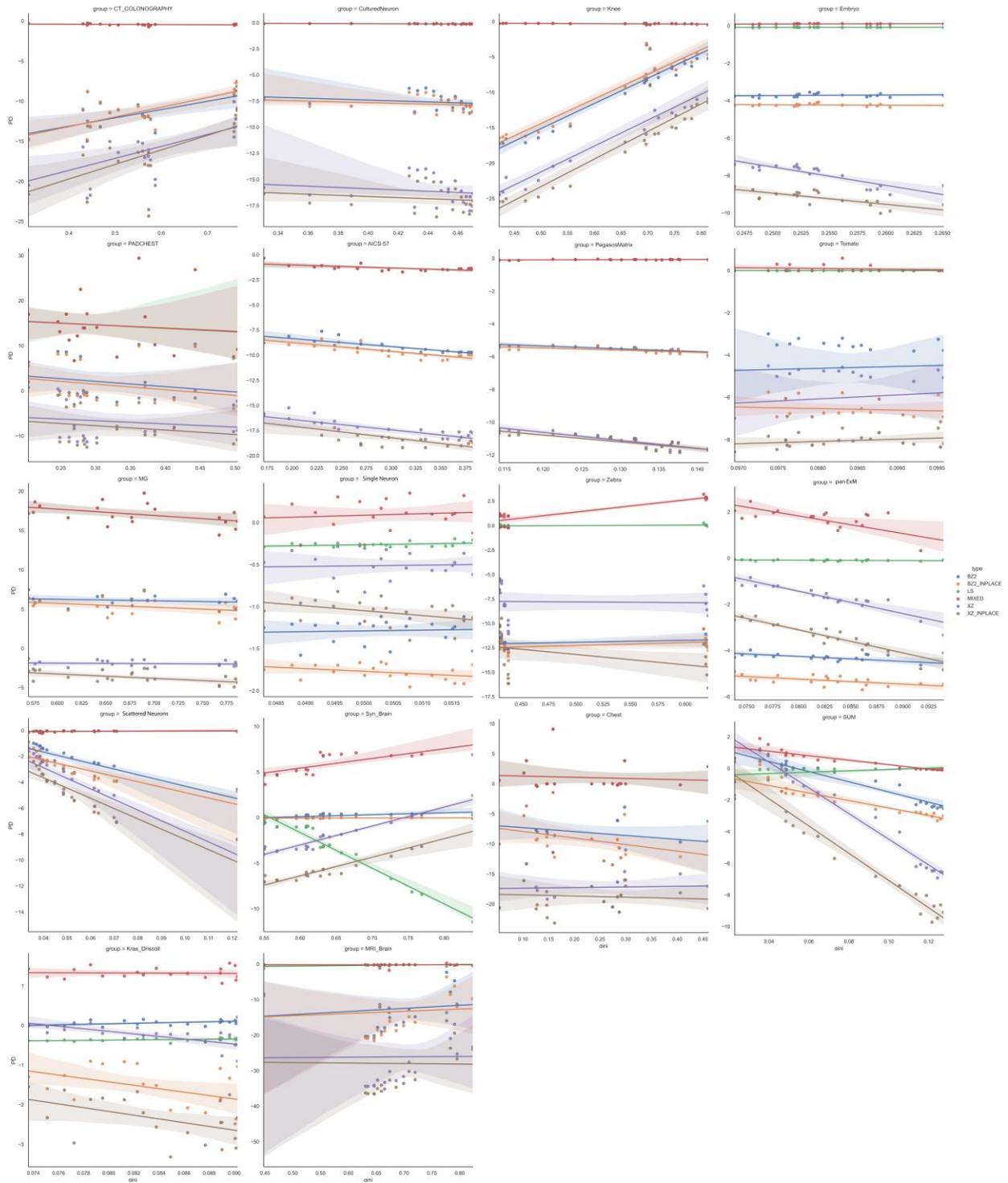

***Supplementary Figure 3:*** *The performance of AGCR with various compression types that integrate existing lossless compression codecs.*

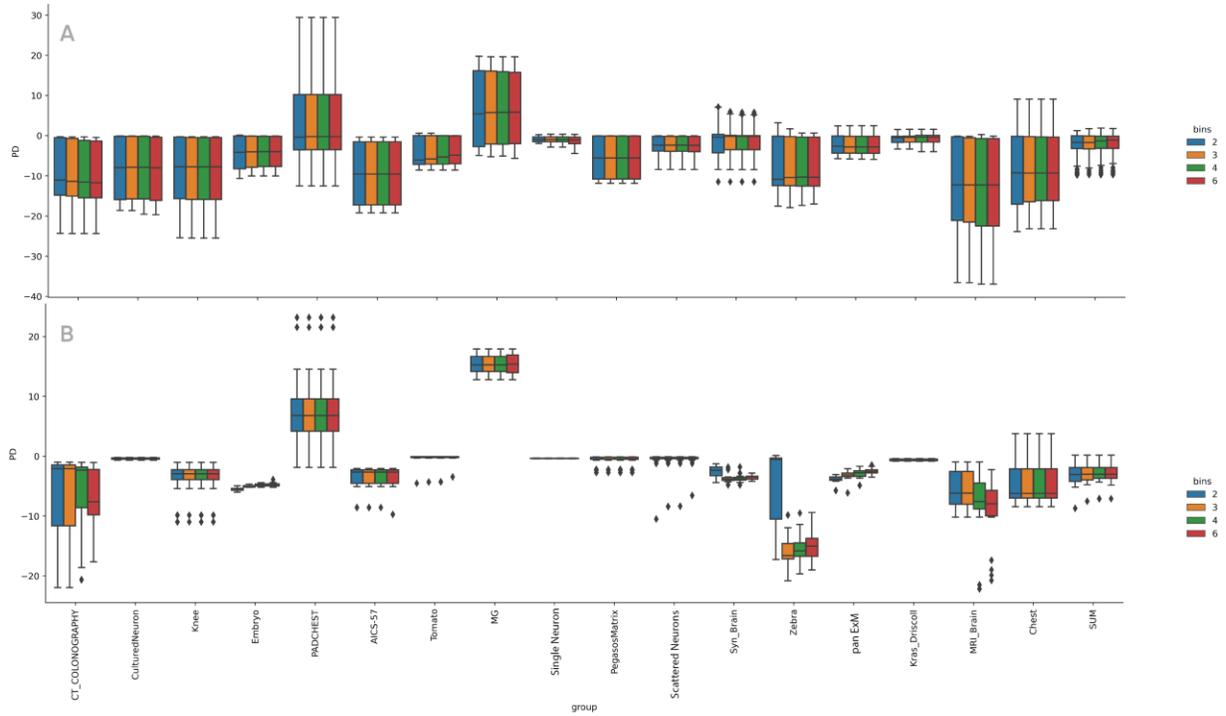

***Supplementary Figure 4:*** *The effect of specifying the number of bins in the automatic (A) configuration of AGCR, and the exact configuration (B). The Gaussian and shape reduction components of (A) reduce the complexity of stored regions and thus choice of bin is not as critical. In (B), AGCR is run with no input template or shape reduction, so 2 bins tends to perform the best due to shape file bloating. In many datasets, our k-level thresholding technique will not support above 2 bins due to the inverse Gini index as the primary determination factor for forming histogram bins. In such scenarios bin choice of 2-6 matches the performance of 2.*

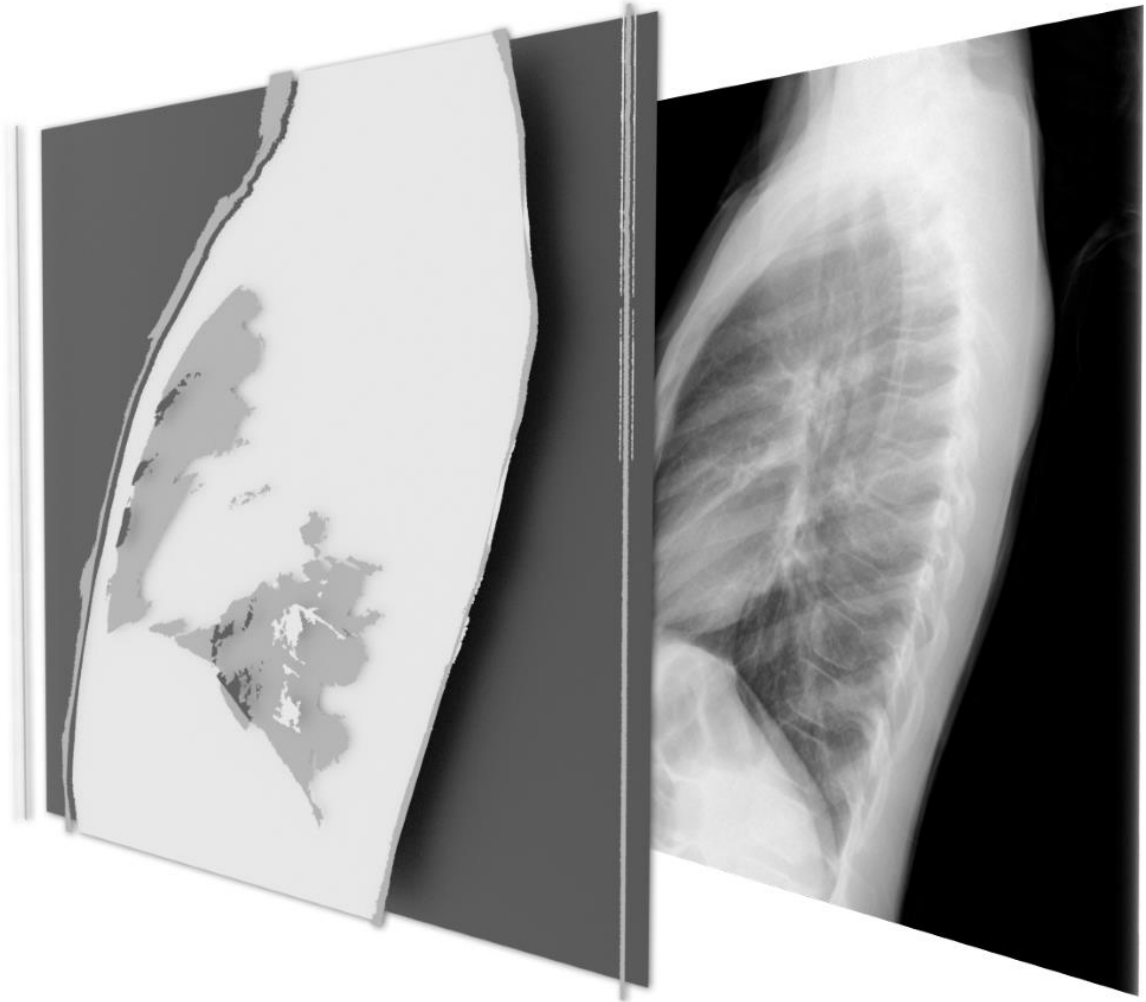

***Supplementary Figure 5:*** *An example visualization using a default Gini-based 3-level thresholding. Manual regions can easily be provided via an alternative threshold pipeline or direct template specification for fine-tuning visualizations.*

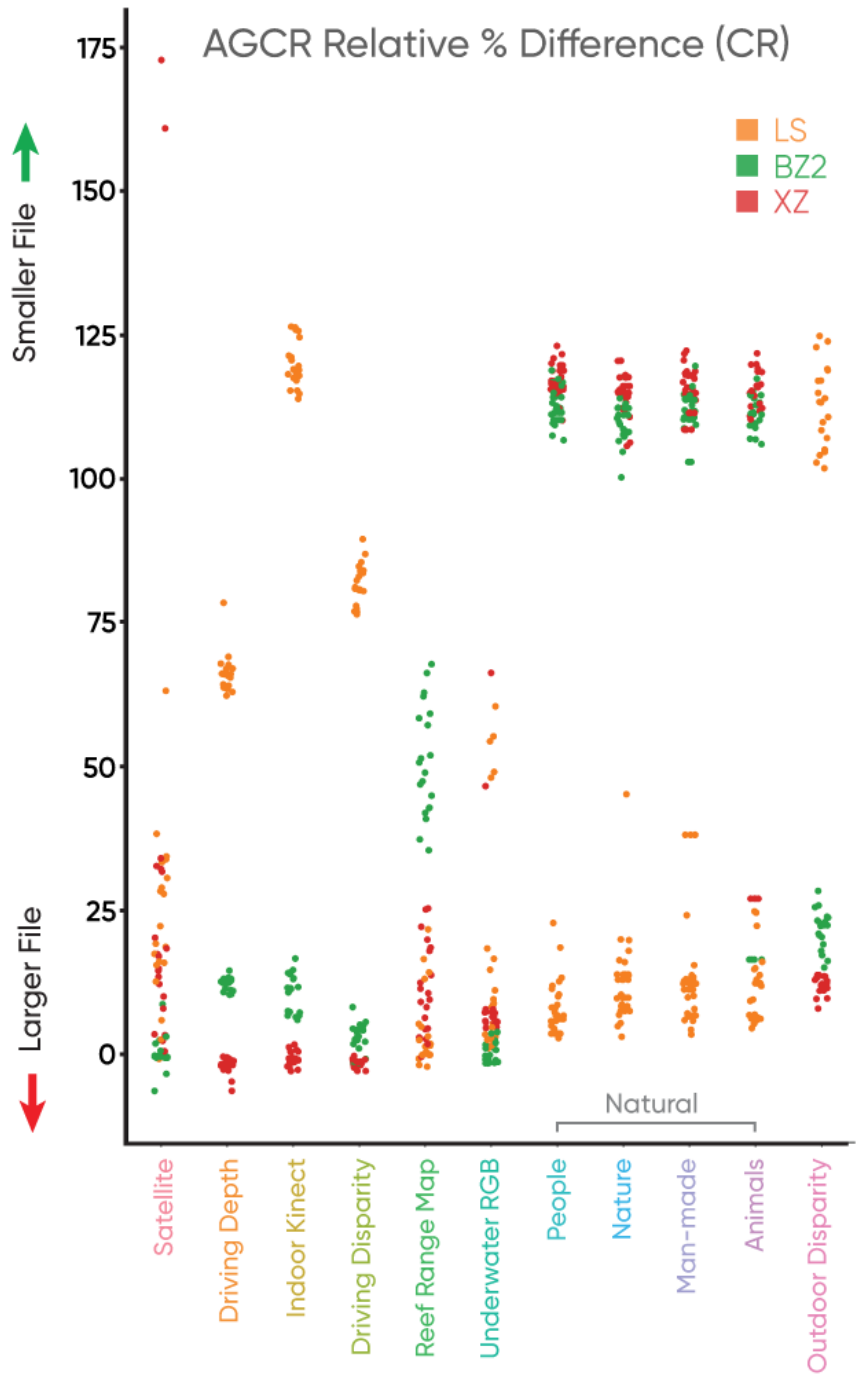

***Supplementary Figure 6:*** *We illustrate relative performance (AGCR vs. alternative) as compared with each compression codec for upwards of 173% improvement over XZ and 126% improvement over JPEG-LS.*